\documentclass{emulateapj}  

\usepackage{graphicx}
\usepackage{natbib}
\usepackage{color}

\newcommand\eq{\begin{equation}}
\newcommand\eeq{\end{equation}}
\newcommand\eqn{\begin{eqnarray}}
\newcommand\eeqn{\end{eqnarray}}

\newcommand\Msol{$M_{\odot}$}

\newcommand\sloani{$i^\prime$}
\newcommand\sloang{$g^\prime$}
\newcommand\sloanr{$r^\prime$}

\shorttitle{Strong Gravitational Lensing in Abell 3827 }
\shortauthors{Carrasco et al. }
\slugcomment{Accepted for publication in The Astrophysical Journal Letters 2010 April 29}

\begin{document}

\title{Strong Gravitational Lensing by the Super-massive cD Galaxy in Abell 3827}

\author{E. R. Carrasco\altaffilmark{1}, P. L. Gomez\altaffilmark{1}, 
T. Verdugo\altaffilmark{2}, H. Lee\altaffilmark{1}, R. Diaz\altaffilmark{1}, M. Bergmann\altaffilmark{1}, 
J. E. H. Turner\altaffilmark{1}, B. W. Miller\altaffilmark{1} and M. J. West\altaffilmark{1,3}}

\altaffiltext{1}{Gemini Observatory, Southern Operations Center, AURA, Casilla 603,
La Serena, Chile}
\altaffiltext{2}{Departamento de F\'{\i}sica y Astronomia, Universidad de Valpara\'{\i}so, 
Avenida Gran Breta\~ na 1111, Valpara\'{\i}so, Chile}
\altaffiltext{3}{Present address: European Southern Observatory, Alonso de
C\' ordova 3107, Vitacura, Casilla 19001, Santiago, Chile}

\begin{abstract}
We have discovered strong gravitational lensing features in the core of the nearby cluster 
Abell~3827 by analyzing Gemini South GMOS images. The most prominent strong lensing  
feature is a highly-magnified, ring-shaped configuration of four images around the central 
cD  galaxy. GMOS spectroscopic analysis puts this source at $z \sim 0.2$. Located $\sim$ 20\arcsec~away 
from the central galaxy is a secondary tangential arc feature which has been identified as a 
background galaxy with $z \sim 0.4$. We have modeled the gravitational potential of the cluster 
core, taking into account the mass from the cluster, the brightest cluster galaxy (BCG) and other galaxies. 
We derive a total 
mass of $(2.7\,\pm\,0.4) \times 10^{13}$ \Msol\ within 37 h$^{-1}$ kpc. This mass is an order of 
magnitude larger than that derived from X-ray observations. The total mass derived from  
lensing data suggests that the BCG in this cluster is perhaps the most massive galaxy in the nearby 
universe.  
\end{abstract}


\section{Introduction} \label{sec:intro}

As the densest galaxy environments known, the cores of massive clusters are
expected to host the strongest dynamical evolution. Such cluster cores are
found to be dominated by early-type, D or cD galaxies that are also the
brightest cluster galaxies (BCGs). These have extended luminous
haloes \citep{schombert1987}, which are not necessarily smooth
\citep{johnstone1991} and contain multiple or complex nuclei
\citep{rood1979}.  They are located close to the peak of the X-ray emission
\citep{jones1984} and near the kinematical centers of their clusters
\citep{quintana1982,quintana2000}.

There has been a long-running debate over the extent to which BCGs can be assembled through continuous merging of 
galaxies in the cluster potential (``galactic cannibalism'') versus early merging during cluster collapse 
\citep[e.g.][]{west1994,dubinski1998}. However, recent work has shown that BCGs are likely  to form via 
``dry'' or dissipationless major mergers of smaller early-type galaxies 
\citep[e.g.][]{vandokkum2005,bell2006,delucia2007,whitaker2008}. These dry mergers are thought to be the primary 
mechanism through which massive galaxies grow from $z\sim1$ to the present
day and continue populating the upper end 
of the mass function without changing the overall mass density of elliptical systems. It has been pointed out that the 
merger rate increases both with the stellar mass of galaxies and with age \citep{liu2009}. Therefore, there is 
a large fraction of major dry mergers in the nearby universe. Indeed, as many as 3.5\% of BCGs show ongoing evidence of
mergers \citep{liu2009}, and about 10\%~- 20\%~of massive galaxies have undergone a dry merger in the last gigayear
\citep{kochfar2009}.  Thus, dry mergers ultimately lead to the formation of very massive and dense BCGs with large mass-to-light ratios.

We report here the discovery of a multi-component BCG that is, to our knowledge, the most massive galaxy ever seen
in the local universe. This is located in the core of Abell 3827, a massive galaxy clusters in Abell's cluster catalog 
(richness class 2 and Bautz-Morgan type I), with an X-ray emission of $L_{X}$(0.1--2.4 keV) $=$ 2.1 $\times$ 10$^{44}$ erg s$^{-1}$.  
This BCG is perhaps the most extreme example of ongoing galaxy cannibalism known: a super-giant elliptical that appears to be in the throes 
of devouring at least four other galaxies.  Evidence for a recent major merger is also supported by the appearance of 
an extended asymmetric halo at the center of the cluster. The super-giant galaxy also shows features arising from strong 
gravitational lensing, the most prominent being a surrounding, magnified ring-shape configuration of four similarly-shaped images. 
The existence of such features provides a unique opportunity to study the mass distribution and the evolution of BCGs with unprecedented 
spatial detail. 

This Letter is organized as follows. In section 2, we summarize our observations and data reduction. In section 3,
we describe our results, including the strong lensing modeling. In section 4 we present further discussion. 
Throughout this Letter we use a standard cosmology of $H_{0}=70$ $h$ km~s$^{-1}$~Mpc$^{-1}$, 
$\Omega_{m}=0.3$ and $\Omega_{\Lambda}=0.7$. At the redshift of Abell 3827, 1\arcsec\ corresponds 
to 1.83~$\!h^{-1}$ kpc.
 
\section{Observations} \label{sec:observation}

All imaging and spectroscopic data were collected with the Gemini Multi-Object Spectrograph \citep[GMOS,][]{hoo04} 
at the Gemini South telescope, in queue mode. 

\begin{figure*}[!htb]
\centering
\includegraphics[width=0.9 \textwidth]{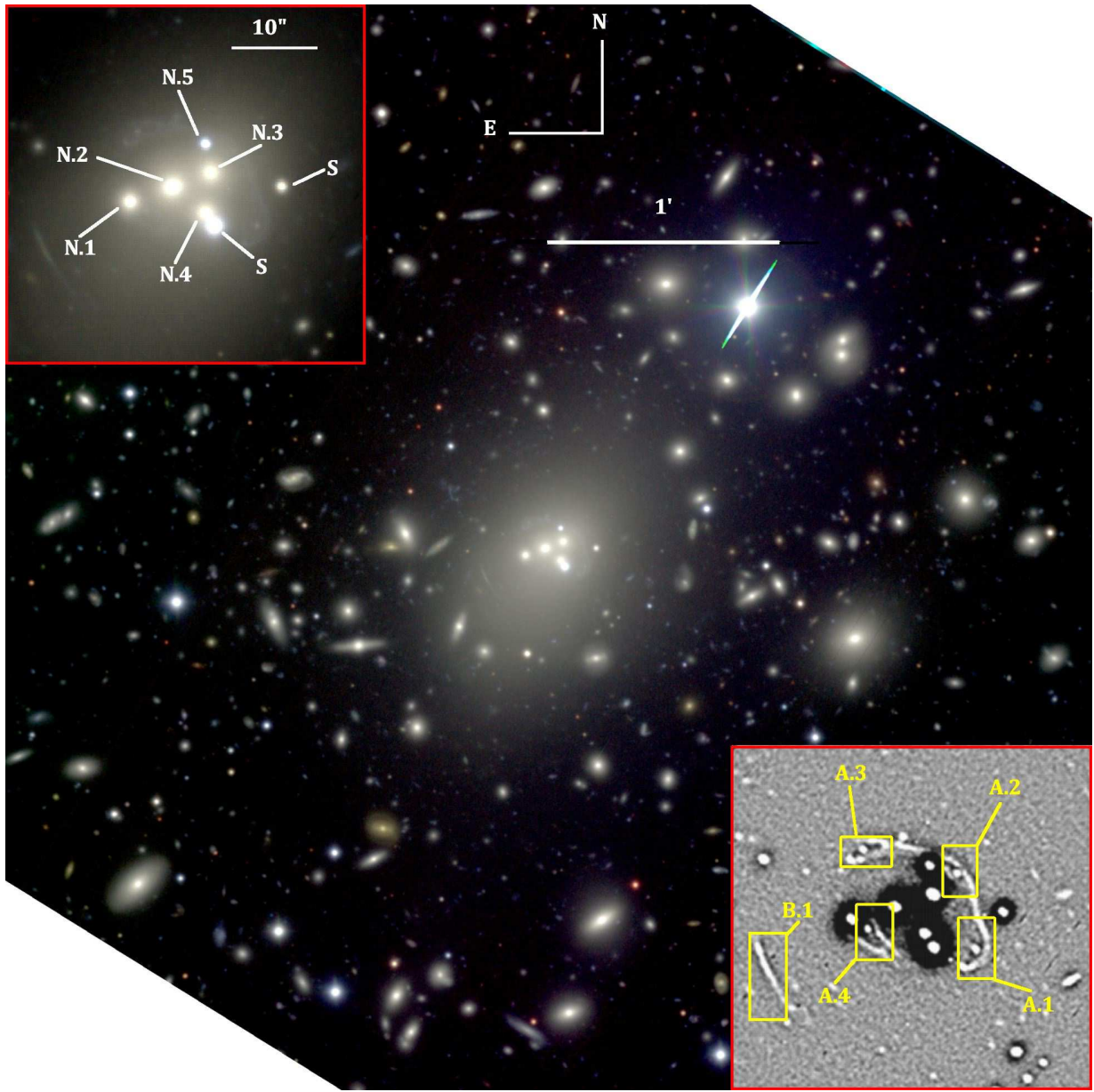}
\caption[]{Color-composite image of Abell 3827 . The field of view is $5\arcmin\!\times\!5\arcmin$ ($\sim 0.55\,\times\,0.55$ 
h$^{-2}$ Mpc$^{2}$ at the distance of the cluster). The cD galaxy and its  asymmetric halo are clearly seen at the center of the image. 
Top - left panel: the field - of - view is approximately $40\arcsec\!\times\!40\arcsec$. The five nuclei contained in the the center are labeled 
N.1 to N.5 and two stars are labeled S. Bottom-right panel:  The field-of-view and orientation are similar to that of the top-left panel. 
The ``central'' arcs are labeled A.1 to A.4 and the tangential arc is labeled B.1. With the central nuclei 
removed and the image display scale chosen to maximize contrast, all of the gravitational-lensing features are more distinct.
\label{colorimage}}
\end{figure*}

The cluster was imaged with the \sloang, \sloanr~and \sloani~filters on
2007 Nov 7 (UT), in photometric conditions, with good seeing (0$\farcs$4 $-$
0$\farcs$6). Observations were processed with the Gemini IRAF package
(version 1.8) in the standard manner. Calibration on the standard
magnitude system was achieved using observations of stars from \citet{lan92}. Object detection and photometry were performed on 
the \sloanr-band image (2007 Nov 7 UT) with the program SExtractor \citep{ber96}. The {\rm MAG$\_$AUTO 
parameter was adopted as the total magnitude for the detected objects. Colors were derived by measuring fluxes 
inside a fixed circular aperture of  1$\farcs$5 ($\sim 10$ pixels) in all filters, corresponding to a physical aperture of 2.75 h$^{-1}$ 
kpc at the distance of the cluster. All objects with  SExtractor {\em stellarity} flag $\le0.8$ were selected as galaxies. We 
estimate that  the catalog is complete down to \sloanr$=$23 mag, since the number counts start to turn over at this value. 
The final catalog contains the total magnitudes, colors and  structural  parameters for 747 galaxies brighter than 
\sloanr$=23$ mag. Figure \ref{colorimage} shows the color-composite \sloang, \sloanr, \sloani~ image of the central 
region of Abell 3827 (0.55 $\times$ 0.55 $h^{-2}$  Mpc$^{2}$).  Five central elliptical galaxies (N.1 to N.5) are
embedded in a common asymmetric halo  (top-left panel). The  asymmetric halo has an ellipsoidal form, with $a\sim$ 1\farcm3 
and $b\sim$ 1\arcmin~ for the semi-major and semi-minor axes, respectively}. Strong lensing features detected around 
the central cD galaxy (at radii $\sim$ 8\arcsec~and $\sim$ 20\arcsec) are visible in the images located in the top-left 
and bottom-right panels of Fig. 1.

\begin{deluxetable*}{cccccrrrccc}
\tabletypesize{\footnotesize}
\tablecaption{Main data of the five central ellipticals galaxies located in the core of Abell 3827\label{tab1}}
\tablenum{1}
\tablecolumns{11}
\tablewidth{0pc}
\tablehead{\colhead{Galaxy} & \colhead{ID} & \colhead{R.A.(2000)} & \colhead{Decl.(2000)} & \colhead{V$_{hel}$} & \colhead{$\delta V_{z}$} &
\colhead{x} & \colhead{y} & \colhead{\sloanr} & \colhead{\sloani} & \colhead{(\sloanr$-$\sloani)$_{1\farcs5}$}\\
\colhead{} & \colhead{} & \colhead{($^{h}$ $^{m}$ $^{s}$)} & \colhead{(\arcdeg~\arcmin~\arcsec)} & \colhead{(km s$^{-1}$)} &
\colhead{(km s$^{-1}$)} & \colhead{(arcsec)} & \colhead{(arcsec)} & \colhead{(mag)} & \colhead{(mag)} & \colhead{(mag)} \\
\colhead{(1)} & \colhead{(2)} & \colhead{(3)} & \colhead{(4)} & \colhead{(5)} & \colhead{(6)} & \colhead{(7)} & \colhead{(8)} &
\colhead{(9)} & \colhead{(10)} & \colhead{(11)}}
\startdata
1323 & N.1 & 22 01 53.99 & $-$59 56 45.14 & 29548 & $-$115 &   13.88 & $-$1.92 & 17.03 & 16.51 & 0.51 \\ 
1324 & N.2 & 22 01 53.31 & $-$59 56 43.14 & 29763 &    100 &    3.75 &    0.08 & 16.19 & 15.67 & 0.52 \\
1325 & N.3 & 22 01 52.69 & $-$59 56 41.22 & 29951 &    288 & $-$5.75 &    2.00 & 16.18 & 15.56 & 0.55 \\
1326 & N.4 & 22 01 52.76 & $-$59 56 46.12 & 28590 &$-$1073 & $-$4.63 & $-$2.91 & 16.23 & 15.81 & 0.43 \\
1327 & N.5 & 22 01 52.79 & $-$59 56 37.73 & 33817 &   4154 & $-$4.13 &    5.48 & 18.08 & 17.70 & 0.35 \\ 
\enddata
\tablecomments{(1) $-$ Galaxy number in the Source Extractor catalog, (2) $-$ Galaxy ID in Fig.1, (3) and (4) $-$ R.A. and decl.
(J2000.0), (5) $-$ heliocentric radial velocity, (6) $-$ difference between the heliocentric radial velocities of each component 
and the systemic velocity, (7) and (8) $-$ coordinates relative to the center of mass of the system  located at $\alpha_{2000}=$ 22$^{h}$ 01$^{m}$ 53$\fs$06, $\delta_{2000}=$ $-$59\arcdeg~43\arcmin~43$\farcs$22. (9) and (10) $-$ total magnitudes, (11) $-$ galaxy colors measured inside a fixed aperture of diameter 1$\farcs$5.}
\end{deluxetable*}

Long-slit spectra of the central five elliptical galaxies were obtained on 2007 Jun 3 (UT) during bright time, 
through thin-cloud cover and $\sim$ 1\arcsec\ seeing. The observations were performed using the nod-and-shuffle 
mode, with the R400 grating centered at 8100\AA; this setup was chosen to minimize the effect of moon 
illumination and to obtain good sky subtraction. To avoid second-order contamination from blue light in the red part of these 
spectra, the OG515 blocking filter was used. In addition, multi-object spectroscopic (MOS) observations of 
galaxies in the Abell 3827 field were obtained on 2007 Sep 7 (UT), during
dark time, in photometric conditions and with seeing 
between 0$\farcs$7 and 0$\farcs$9. Two masks were observed with 0$\farcs$75 slit widths, using the B600 grating 
centered at 5220\AA. During the observations, all spectra were dithered to
account for the gaps in the GMOS CCDs. Spectra were reduced and calibrated
in the standard manner for each mode.

\section{Results} \label{sec:res}

We have measured radial velocities for 67 galaxies  using cross-correlation techniques or emission-line fitting. Of these 67
galaxies, 56 are within $\pm\,2500$ km~s$^{-1}$ of the cluster redshift. Using the  bi-weight estimator 
for location and scale \citep{beers90}, we calculate an average velocity for the cluster of 29663$\,\pm\,$155 km~s$^{-1}$
and a line-of-sight velocity dispersion of 1142$\,\pm\,$125 km~s$^{-1}$, with 55 member galaxies.  The virial mass inside a radius 
of  0.3 h$^{-1}$ Mpc is $(3.86_{-0.28}^{+0.46}) \times 10^{14}\,h^{-1}$ \Msol, with uncertainties at the 68\% confidence intervals.  

In the cluster core, we confirm membership for four of the central
elliptical galaxies. There are also two stars in the foreground (see Fig. 1).  Three of the  core galaxies have similar colors (see Table 1) and their velocities 
are consistent with the average cluster velocity (within $\sim$ 300 km~s$^{-1}$). Galaxy N.4 is slightly  bluer than N.1, N.2 and N.3 and has a  measured radial velocity $\sim$ 1000 km s$^{-1}$ higher than the systemic  velocity. 
Galaxy N.5 may not be in the cluster, but the background.

\subsection{Strongly lensed features in the cluster core} \label{sec:arcs}

Thanks to the superb image quality delivered by GMOS and the Gemini South telescope, we have been able to detect several 
strongly-lensed features around the central cD galaxy in the core of the cluster.  A thin, extended tangential arc (B.1 in Fig. 1) 
is seen  at $\sim$  20\arcsec~south-east from the central galaxy, with apparent length $\sim\,11\arcsec$ and thickness 
$\sim\,0\farcs7$. Surrounding the central core galaxies is a second and more prominent arc feature---a highly-magnified, 
ring-shaped configuration of four images around the central cD galaxy, with a radius of $\sim$~8\arcsec. Only three of the arc 
components are visible in the inset color image in Fig. 1 (A.1, A.2 and A.3).  The fourth image (A.4 in Fig. 1) is between galaxies N.1 
and N.2 and is obscured by the halo of the cD galaxy.  All four features can be seen in detail in the bottom-right panel in Fig.~1 (see 
the caption for details). These features are very reminiscent of the lensed images seen in Cl0024+1654 \citep{colley1996}, although 
the source galaxy in the present case is closer than $z\sim1$ (see below). 

The redshifts of the lensed sources were determined from long-slit
observations taken on 2007 Nov 15-16,
using the R150 grating centered at 7150\AA. For the feature with the ring-shaped configuration (system A), 
we derived a redshift for one of the images located north-west of the
cluster core (A.2). Two emission lines 
($H\alpha$ and [N~II]$ \lambda\,6583$\AA) are seen in the spectrum for A.2;
unfortunately no other lines were detected, due to light contamination from
the cD halo. From the two emission lines, the estimated redshift for A.2 is $z = 0.20443\,\pm\,0.00073$.

The symmetric distribution of images in system A and the ring-shaped
structure connecting arcs A.1, A.2 and A.3 (and possibly A.4) suggest that
the images are of the same source galaxy. Therefore, we assume that all
four images come from the same source and have the same redshift. This
assumption should be taken with caution, without spectroscopy of the
remaining system A images for confirmation, but is supported by our
modeling, discussed in the next section.

The subtraction of light from the halo of the cD galaxy, as well as
subtraction of all galaxies and the stars from the cluster core, highlights
the ring-shaped structure connecting arcs A.1, A.2, A.3 and possibly A.4
(see Fig. 1). The existence of this structure only $\sim 15\,h^{-1}$ kpc
from the center of the cD galaxy gives us a unique opportunity to study the
dynamics and mass distribution of a large BCG at unprecedentedly small
scales.

Using several emission-lines ([OII] $\lambda\,3727$\AA, $H\!\beta$, [OIII]
$\lambda\lambda\,4959,5007$\AA) seen in the spectrum of the south-east
tangential arc, B.1, we derived a redshift of $z=0.40825\,\pm\,0.00072$.
The relatively long, thin appearance of this arc suggests that the source
galaxy lies very close to the inner fold caustic, assuming the lens cluster
has an elliptical spatial distribution of mass
\citep[e.g. see][]{fort1994,narayan1999}. If true, there must be three
fainter counter-images: one to the north-east, one to the south-west and
one at the center of the cluster core.

\subsection{Strong Lensing modeling and mass determination} \label{sec:mass}

To produce the features described in the previous section, the cluster core must be very massive. In order to 
reconstruct the mass distribution in the core (the region enclosed by system A), we used the parametric model 
implemented in the publicly available {\em LENSTOOL}\footnote{http://www.oamp.fr/cosmology/lenstool/} ray-tracing 
code \citep{jul07}. To model the lens, we used two clumps of different scales---a large-scale halo, representing both 
the matter inside the giant cD galaxy and the dark matter halo of the cluster, and four smaller-scale
clumps, modeling the small-scale perturbations associated with the galaxies embedded in the cD galaxy. 
For all components we adopt a dual Pseudo-Isothermal Elliptical Mass Distribution (dPIEMD; see
\citealp{eliasdottir2007}). The dPIEMD can be characterized by seven
parameters: the center position $(X,Y)$, the ellipticity $\epsilon$, the
position angle $\theta$ and the parameters of the density profile: the
velocity dispersion $\sigma_0$ and two characteristic radii $r_{core}$ and
$r_{cut}$. For the large halo, we left most parameters free, with broad uniform priors.   
Only $r_{cut}$ remained fixed at a value of 1500~h$^{-1}$ kpc.  In the small-scale clumps, the parameters of the 
density profiles were scaled as a  function of their galaxy luminosities
\citep[see ][ and references therin]{jul07}. Using as a scaling factor the luminosity $L^{*}$, associated with the
\sloang-magnitude of central galaxy N.3 (see Fig. 1 and Table 1), we searched
for the values of $\sigma_0^{*}$ and $r_{cut}^{*}$ that yield the best fit,
fixing $r_{core}^{*}$ at 0.15h$^{-1}$ kpc. The best-fitting parameters of
the model are listed in Table 2.

\begin{deluxetable}{lcccccccc}
\tablecaption{Best-fitting parameters\label{tab2}}
\tabletypesize{\footnotesize}
\tablenum{2}
\tablecolumns{9}
\tablewidth{0pt}
\tablehead{ \colhead{Comp.}    &  \multicolumn{1}{c}{X} & \multicolumn{1}{c}{Y}  &  \colhead{$\epsilon$} &
\multicolumn{1}{c}{$\theta$} & \multicolumn{1}{c}{$r_{core}$} & \multicolumn{1}{c}{$r_{cut}$} & \multicolumn{1}{c}{$\sigma_0$}  \\
\colhead{} & \colhead{$('')$} & \colhead{$('')$} & \colhead{ } & \colhead{$(^{\circ})$} & \colhead{(kpc)} & \colhead{(kpc)} &
\colhead{(km s$^{-1})$}  }
\startdata
Cluster+cD     & $-$0.9    & $-$2.5  & 0.14   & 56                      & 32 & [1500]                        & 1174             \\
N.3($L^{*}$)    &  [2.77]   &  [2.82]  & [0]   & [0]   & [0.15] & 8                  & 256              \\
\enddata
\tablecomments{ Coordinates are relative to the center of image 2 ($\alpha_{2000} = $22$^h$ 01$^m$ 53$\fs$15, $\delta_{2000}$ = -59$\degr$ 56$\arcmin$ 43$\farcs$31). Values in square brackets are not optimized.}
\end{deluxetable}

\begin{figure}[!htb]
\centering
\includegraphics[width=0.95 \columnwidth]{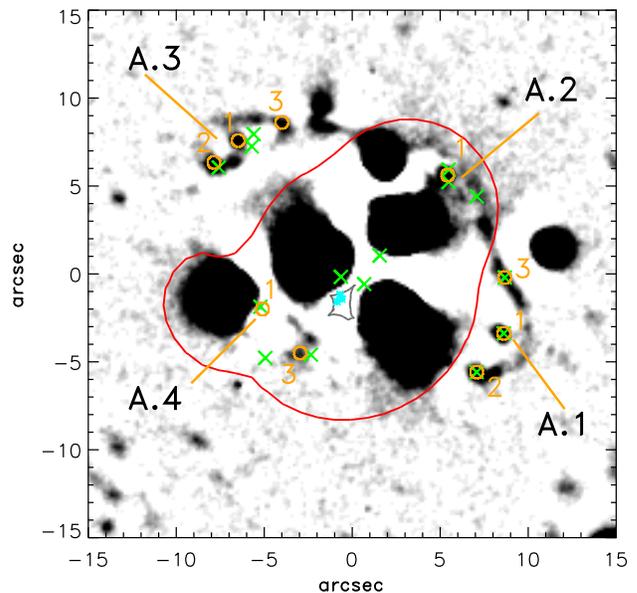}%
\caption[]{\sloang - band image of Abell 3827  with the local median average subtracted. North is up and east is left. The external 
critical line (red) and the associated caustic line (gray) for a source at z = 0.2 are depicted. Orange circles are the measured 
image positions. Green crosses represent the predicted image positions, using as input the three substructures that constitute
arc A.1. We also show with small asterisks (cyan) the positions in the source plane associated with these three points.  For clarity, 
we do not depict the demagnified images produced by strong galaxy - galaxy lensing events.
\label{model}}
\end{figure}

Because the ring structure shows multiple subcomponents, we selected the most reliable set of multiple images and 
associated them with three different background sources in order to make our fit. Considering an uncertainty in the 
position of any image equal to 0$\farcs$2, we found after the optimization in the image plane a 
$\chi_{DOF}^{2}$ = $\chi^2/DOF = 69/4$ $\sim$ 17. Our model reproduces well the positions of the observed 
subcomponents, with a mean scatter less than 0$\farcs$8. To highlight this, we show in Figure~\ref{model}, the model-predicted 
counter-images of the substructures in arc A.1 (green crosses), which are in agreement with the image positions for all
arcs (orange circles). The fit also predicts a central demagnified image which is lost in the cD light distribution.  It is important 
to note that our model is oversimplified, since we are assuming spherical halos for the galaxies. A detailed model of Abell~3827 
is beyond the scope of the present work.

Given our best model (see Table 2), we calculated the total mass inside a radius of  20\arcsec (location of the 
tangential arc B.1 at $\sim$ 37 h$^{-1}$ kpc) and found $M = (2.7\pm0.4) \times 10^{13}$ \Msol.  This mass is 
slightly greater than that enclosed in Abell~1689 within a similar radius (see Fig.\ 6 in \citealp{lim07}). We estimated 
the mass of the cD galaxy as being $\sim$ 30$\%$ $-$ 50$\%$ of the total mass of the cluster within 20\arcsec~
\citep{lim07}. This implies a cD galaxy mass between  $8.1 \times 10^{12}$ M$_{\sun}$ and  $1.3 \times 10^{13}$ M$_{\sun}$. 
Even assuming a conservative value of  30$\%$, this would mean that the central cD galaxy in Abell 3827 is perhaps the most 
massive galaxy observed in the local universe. 

\section{Discussion} \label{sec:discussion}

Using the strong-lensing model presented above, we have analyzed the mass distribution to unprecedented spatial 
resolution. The cluster core is spatially very concentrated. Even if we assume a conservative fraction for the 
mass of the central cD galaxy (see above), this galaxy is very massive. The cD galaxy could be an extreme example 
of the effects of dry mergers on the mass of BCGs; dry mergers can produce an increase (by a factor of up to
3) in the dark-matter-to-stellar mass ratio for the most massive systems at present \citep{ruszkowski2009}.

\begin{figure}[!htb]
\centering
\includegraphics[width=0.95 \columnwidth]{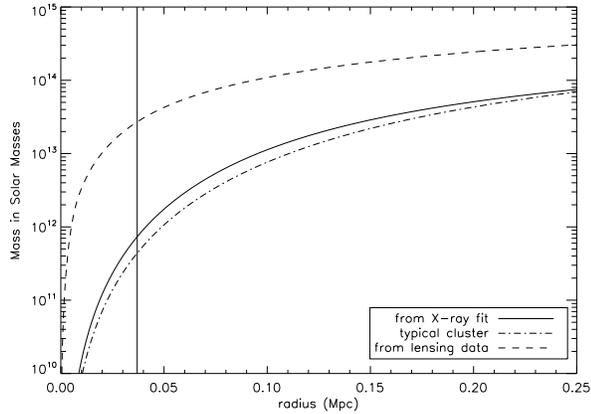}
\caption[]{Mass profile derived from the best fit model for the X-ray gas distribution (continuous line).
The dash-dotted line shows the profile of the X-ray distribution for a more typical cluster. The 
short-dashed line shows the mass profile derived from strong lensing. The vertical solid line shows the 
location of the tangential arc (B.1) at 37 h$^{-1}$ kpc.
\label{mass}}
\end{figure}

We have compared the total mass derived from our strong-lensing analysis with that derived from the X-ray 
gas. We have calculated the total  mass as a function of radius for Abell 3827, based on a recent X-ray analysis 
by L. Valkonen et al. (2010, private communication) . We derive the total cluster mass by assuming hydrostatic 
equilibrium and using their best fit for the three-dimensional distribution of X-ray emitting gas. L. Valkonen et 
al. (2010, private communication) report 
$M_{200} = 9.7 \times 10^{14}$ M$_{\sun}$ inside a radius $r_{200}=2.84$ h$^{-1}$ Mpc, derived from the 
$M-T$ relation by \citet{salhen2009} for a cluster temperature of $kT = (7.15 \pm 0.18)$ keV.  Fig.~\ref{mass} 
shows the radial distribution of the total mass derived from the X-ray gas and from the 
strong-lensing analysis.  For the X-ray gas, we plot mass profiles: (1) for a typical cluster with $r_{core} = 250$ 
h$^{-1}$ kpc and $\beta = 0.7$ (dash-dot line), and (2) for the derived total mass of Abell 3827 within 37 
h$^{-1}$ kpc (solid vertical line). Finally, we plot the mass profile for Abell 3827 derived from strong lensing 
as a dashed line. At all radii, the mass derived from strong lensing is at least a factor of 10 larger than that
derived from X-ray data. Discrepancies between strong-lensing cluster masses and X-ray cluster masses 
for the same clusters have been reported previously in the literature \citep[e.g.][]{gitti2007,halkola2008}.
Moreover, significant biases of up to $\sim 50$\% may be introduced into strong-lensing mass estimates when models are 
extrapolated outside the Einstein ring \citep{meneghetti2009}.  Analysis using both strong-lensing and X-ray methods at 
small radii show that differences can grow by as much as an order of magnitude \citep{verdugo2007}.

How can we explain this discrepancy? For the lower limit on mass, within $3\sigma$ error of the strong-lensing
fit, we obtain $6.3 \times 10^{12}$ M$_{\sun}$ inside the ring structure of radius 10\arcsec~($\sim~19$~h$^{-1}$ kpc).  
Hence, the most likely uncertainties, related to extrapolation of the strong-lensing model to larger 
radii, are unable to explain our factor-of-ten discrepancy.  Assuming a spherical mass distribution, the mass inside  
system A is $(9.6 \pm 0.9) \times 10^{12}$ M$_{\sun}$. The large difference between the strong-lensing mass 
and X-ray mass could be related to the dynamical state of the cluster, to the mass inferred from the X-ray gas, or 
to both effects. Total X-ray cluster masses derived from flat-core density profiles may be underestimated
by at least a factor of 2 within the central $\sim$ 30 h$^{-1}$ kpc \citep{voigt2006}. Simulations 
\citep{meneghetti2009} of strong-lensing models have shown that physical substructures along the line of sight to 
the cluster can overestimate the derived total mass by a factor of 2; in fact, our GMOS spectra reveal a
bi-modality in the velocity distribution of cluster galaxies, suggesting that Abell 3827 is presently merging.  In all, 
these effects can produce a factor of about 4, still insufficient to explain the large discrepancy between strong-lensing and 
X-ray derived masses. Moreover, if we take out the factor of 4 due to the two effects described above, we are left with a 
cD galaxy with a mass of  $\sim 2 \times 10^{12}$ M$_{\sun}$,  which appears to be one of the most massive galaxies 
known in the local universe. Detailed studies of the central objects at different wavelengths will reveal insights
into the origins and subsequent evolution of this peculiarly massive galaxy.

\acknowledgments

We thank the anonymous referee for insightful comments and suggestions. T.V. acknowledges support 
from FONDECYT (grant 3090025). This work is based on  observations obtained at the Gemini Observatory which is operated 
by the Association of Universities for Research in Astronomy, Inc., under a  cooperative agreement with the NSF on 
behalf of the Gemini partnership: the National Science Foundation (United States), the Science and Technology  Facilities 
Council (United Kingdom), the National Research Council (Canada),  CONICYT (Chile), the Australian Research Council (Australia),
Minist\' erio da  Ci\^ encia e Tecnologia (Brazil) and Ministerio de Ciencia, Tecnolog\'{\i}a e  Innovaci\' on Productiva  (Argentina).  
Program ID: GS-2007A-DD-5, GS-2007B-Q-13.

{\it Facilities:} \facility{Gemini:South (GMOS-S)}

\end{document}